\documentclass[%
 reprint,
 amsmath,amssymb,
 aps,
pra,
]{revtex4-1}

\usepackage{graphicx}% Include figure files
\usepackage{dcolumn}% Align table columns on decimal point
\usepackage{bm}% bold math
\begin{document}

\preprint{APS/123-QED}

\title{Doppler-free spectroscopy of metastable Sr atoms using a hollow cathode lamp}% Force line breaks with \\
%\thanks{A footnote to the article title}%

\author{Yusuke Hayakawa}
\email{yusukehayakawa49@gmail.com}
\author{Takumi Sato}
\author{Chika Watanabe}
\author{Takatoshi Aoki}
\author{Yoshio Torii}
\email{ytorii@phys.c.u-tokyo.ac.jp}

\affiliation{Institute of Physics, The University of Tokyo, 3-8-1 Komaba, Megro-ku, Tokyo 153-8902, Japan}%Lines break automatically or can be forced with \\

\date{\today}% It is always \today, today,
             %  but any date may be explicitly specified

\begin{abstract}
We report on the demonstration of Doppler-free spectroscopy of metastable Sr atoms using a hollow cathode lamp (HCL).  We employed a custom Sr HCL which is filled with a mixture of 0.5-Torr Ne and 0.5-Torr Xe as a buffer gas to suppress velocity changing collisions and increase the populations in all the $(5s5p){}^3P_J(J=0,\ 1,\ 2)$ metastable states. We performed frequency-modulation spectroscopy for the $(5s5p){}^3P_0-(5s6s){}^3S_1$, $(5s5p){}^3P_1-(5s6s){}^3S_1$, $(5s5p){}^3P_2-(5s5d){}^3D_2$, and $(5s5p){}^3P_2-(5s5d){}^3D_3$ transitions with sufficient signal to noise ratios for laser frequency stabilization.	We also observed the hyperfine transitions of $(5s5p){}^3P_2-(5s5d){}^3D_3$ of $^{87}\mathrm{Sr}$ . This method would greatly facilitate laser cooling of Sr.
\end{abstract}

\maketitle

\section{Introduction}

The optical lattice clock \cite{opticallatticeclock} has reached an uncertainty of $10^{-18}$ \cite{OLC10-18/2, OLC10-18/3}, which is three orders of magnitude better than a conventional microwave atomic clock, and is one of the most promising candidates for the redefinition of a second. The optical lattice clock has the potential to detect the gravitational potential difference of 1 cm on a time scale of 1 second \cite{KatoriOLC}, and is opening the way to the new field of relativistic geodesy \cite{geodesy}: the application for chronometric leveling has recently been explored using a transportable optical lattice clock \cite{transportableolc}. From this perspective, the portability of optical lattice clocks is crucial and facilitation of laser cooling technique is highly desirable.

\begin{figure}[ttt]
	\begin{center}
		\includegraphics[width=75mm]{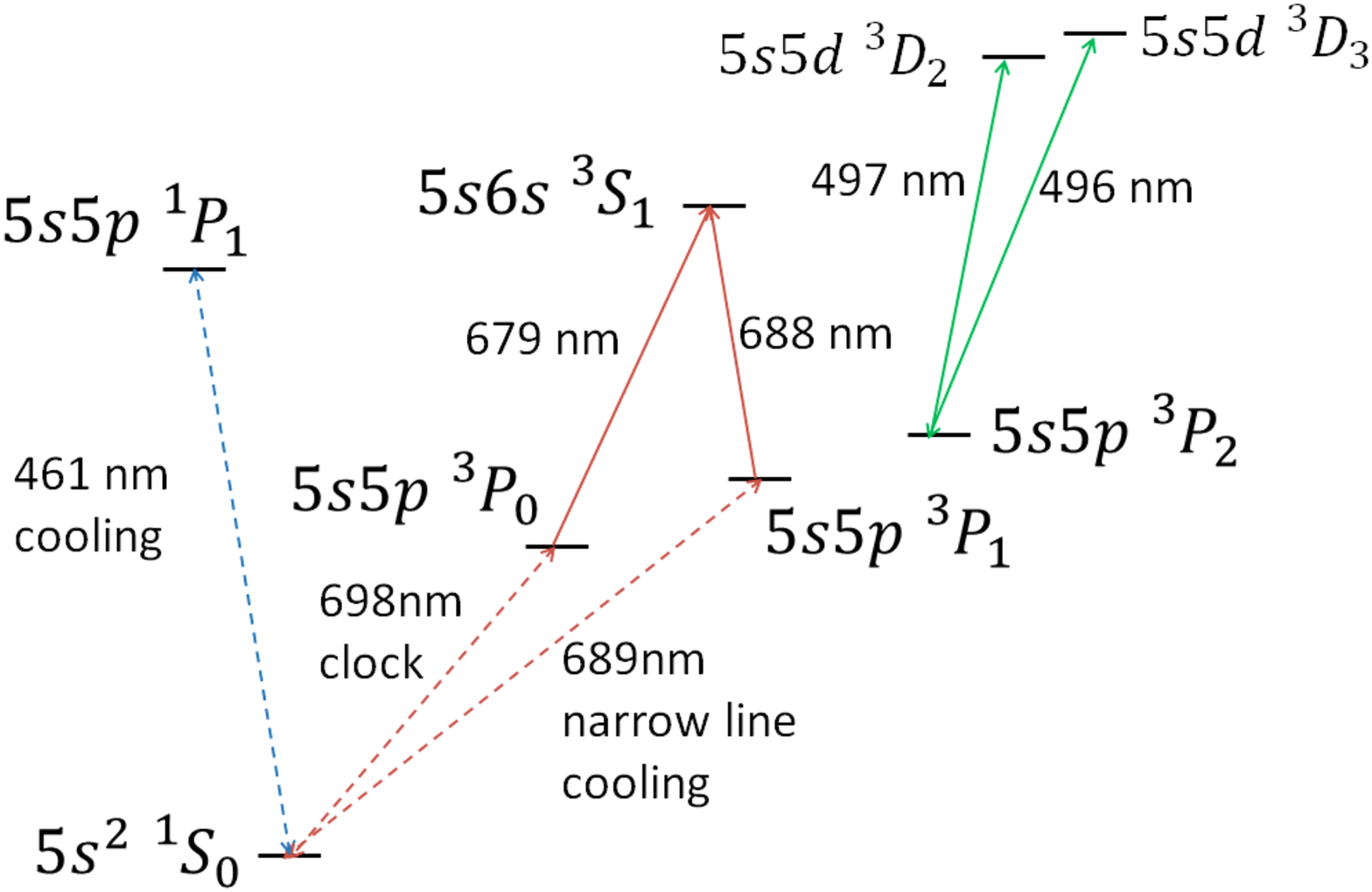}
		\caption{Relevant energy levels of Sr. The transitions that are indicated by solid lines are investigated using hollow cathode lamps in this work. Cooling and clock transitions are indicated by dashed lines.}
		\label{fig:energylevelofsr}
	\end{center}
\end{figure}

Laser frequency stabilization for the cooling transition ($(5s^2) {}^1S_0-(5s5p){}^1P_1$, see Fig. \ref{fig:energylevelofsr}) of Sr can be easily performed by direct Doppler-free spectroscopy using a hollow cathode lamp (HCL). The HCL provides Sr atoms in the ground state ($(5s^2) {}^1S_0$) based on a sputtering process \cite{461hcl, simple461nmlasersystem}. The optical transitions including the metastable $(5s5p){^3P_J\ (J=0,\ 1,\ 2)}$ states as the lower level are also important for repumping \cite{Srrepump, Srrepump2, Srrepump3, Srrepump4, Srrepump5, Srrepump6, energylevel3}. A quantum computation scheme based on the transitions including the metastable ${}^3P_0$ and ${}^3P_2$ states has been proposed \cite{QuantumComputation}.  Laser cooling of Sr to quantum degeneracy with the aid of the $(5s5p){}^3P_1-(5s6s){}^3S_1$ (688 nm) transition has been realized \cite{688app}. However, so far, the laser frequencies have been indirectly stabilized to these transitions using a transfer cavity \cite{transfercavity, transfercavity2}, or an accurate wavelength meter \cite{wavelengthmeter}, which make the system complex and expensive.

Recently, Norcia and Thompson demonstrated direct Doppler-free spectroscopy for the $(5s5p){}^3P_2-(5s6s){}^3S_1$ (707 nm) transition using a commercial see-through HCL (Hamamatsu L2783-38Ne-Sr) filled with a Ne gas of 5-10 Torr \cite{accurate} and succeeded in stabilizing the laser frequency to this transition \cite{707}. They also observed error signals for the $(5s5p){}^3P_0-(5s6s){}^3S_1$ (679 nm) and $(5s5p){}^3P_1-(5s6s){}^3S_1$ (688 nm) transitions. However, the signal-to-noise ratios were not sufficient for laser frequency stabilization, mainly owing to velocity-changing collisions (VCCs) and insufficient populations in the $(5s5p){}^3P_0$ and $(5s5p){}^3P_1$ states.

In this paper, we report on a solution to this problem: we employed a custom HCL with a lower buffer-gas pressure (0.5-Torr Ne and 0.5-Torr Xe). We realized direct Doppler-free spectroscopy of the $(5s5p){}^3P_0-(5s6s){}^3S_1$ (679 nm), $(5s5p){}^3P_1-(5s6s){}^3S_1$ (688 nm), $(5s5p){}^3P_2-(5s5d){}^3D_2$ (497 nm), and $(5s5p){}^3P_2-(5s5d){}^3D_3$ (496 nm) transitions  (indicated by solid lines in Fig. \ref{fig:energylevelofsr}) with sufficient signal-to-noise ratios for laser frequency stabilization. In the Doppler-free spectrum of the $(5s5p){}^3P_2-(5s5d){}^3D_3$ transition, we observed eight of the fifteen hyperfine transitions of fermionic $^{87}\mathrm{Sr}$, six of which were observed for the first time, and confirmed the hyperfine constants which were recently reported by Stellmer and Schreck \cite{energylevel3}.

\section{experimental setup}

\begin{figure}[t]
	\begin{center}
		\includegraphics[width=75mm]{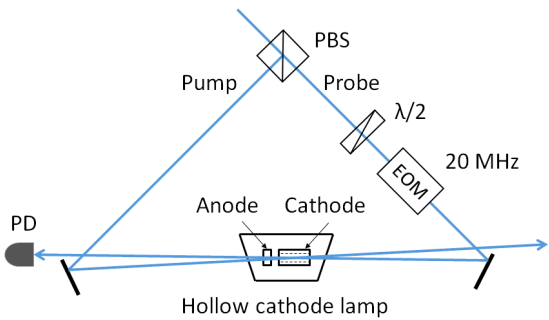}
		\caption{Schematic diagram of the experimental setup. EOM : electro-optic modulator, $\lambda/2$ : half wave plate, PD : photo detector, PBS : polarization beam splitter.}
		\label{fig:expsetup}
	\end{center}
\end{figure}

 An external cavity diode laser (ECDL) was used as a light source for spectroscopy of the ${}^3P_1-{}^3S_1$ (688 nm) and the ${}^3P_0-{}^3S_1$ transitions (679 nm). For spectroscopy of the ${}^3P_2-{}^3D_2$ (497 nm) and the ${}^3P_2-{}^3D_3$ (496 nm) transitions, we also used an infrared ECDL whose frequency was doubled by a waveguide periodically-poled lithium niobate (PPLN) \cite{Akamatsu:11}. Each beam was coupled to a single-mode fiber and delivered to the setup for spectroscopy.

We performed Doppler-free frequency modulation spectroscopy (FMS) \cite{FMS1,FMS2} as shown in Fig. \ref{fig:expsetup}. The pump and probe beams, both of which had 1/e diameters of 320 $\mu \mathrm{m}$, were vertically polarized and intersected with each other in the center of the HCL at an angle of 0.01 rad. The power of the pump (probe) beam was set to 0.3 (0.1) mW. The laser frequency was swept at a rate of 2 GHz/s. For FMS, the frequency of the probe beam was modulated by an electro-optic modulator (EOM) at a modulation frequency of 20 MHz and with a modulation index of 1.4. The probe beam was detected by an amplified photo-detector (Thorlabs PDA10A-EC) and the output signal was fed to a lock-in amplifier (Stanford Research Systems SR44) with a time constant of 1 ms.

We used a custom (0.5-Torr Ne and 0.5-Torr Xe) and a commercial HCL. Both HCLs had the same geometry: the anode was ring-shaped and the cathode had a length of about 20 mm and a bore diameter of 3 mm (Fig. \ref{fig:expsetup}). We typically operated the custom (commercial) HCL with a voltage of 280 (210) V with a discharge current of approximately 15 mA for both HCLs. The reason for choosing a mixed gas of 0.5-Torr Ne and 0.5-Torr Xe was to reduce the total pressure while sustaining a glow discharge at a relatively low voltage (a pure Ne buffer gas requires 1000 V or more to sustain a glow discharge when the pressure is less than 3 Torr). Metastable Sr atoms were mainly created in the center of the hollow cathode, which is called the negative glow region \cite{hollowcathode}. However, as the mean free path of Sr atoms was of the order of 10 $\mu$m  in the commercial HCL filled with a Ne gas of 5-10 Torr, the population of sputtered Sr atoms was small at the center of the hollow cathode \cite{simple461nmlasersystem}. It is expected that a reduced buffer gas pressure will make the mean free path of Sr atoms longer and eventually increase the population of Sr atoms in the metastable state.

\section{Results}

\begin{figure}[t]
	\begin{center}
		\includegraphics[width=75mm]{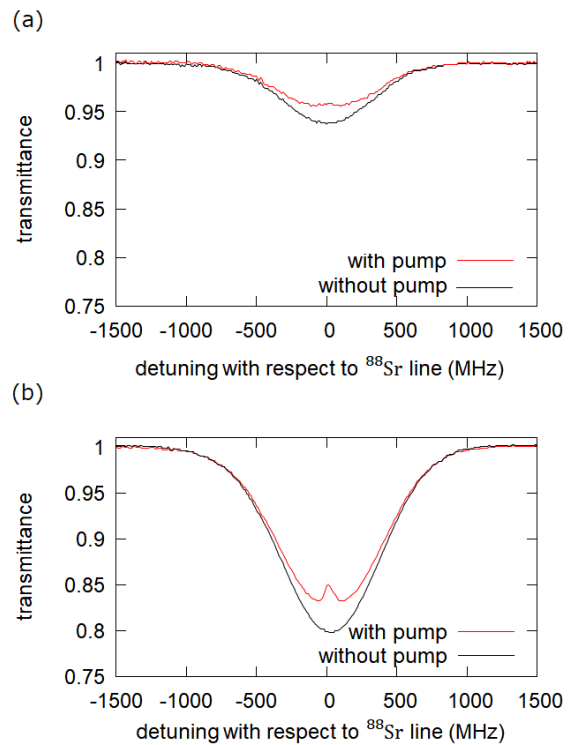}
		\caption{Saturated absorption spectroscopy signals for the $(5s5p){}^3P_2-(5s5d){}^3D_3$ (496 nm) transition using the commercial (a) and custom (b) HCLs. The black (red) traces are taken with (without) the pump beam.}
		\label{fig:496_1}
	\end{center}
\end{figure}

\begin{figure}[t]
	\begin{center}
		\includegraphics[width=80mm]{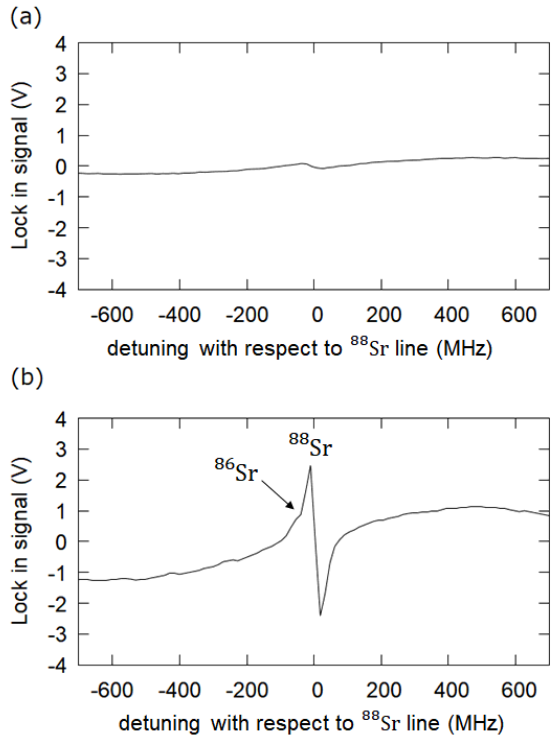}
		\caption{Frequency modulation spectroscopy signals for the $(5s5p){}^3P_2-(5s5d){}^3D_3$ (496 nm) transition using the commercial (a) and custom (b) HCLs. The small dispersive signals in the wings in (b) arise from fermionic ${}^{87}\mathrm{Sr}$ (see text).}
		\label{fig:496_3}
	\end{center}
\end{figure}

\begin{figure}[t]
	\begin{center}
		\includegraphics[width=75mm]{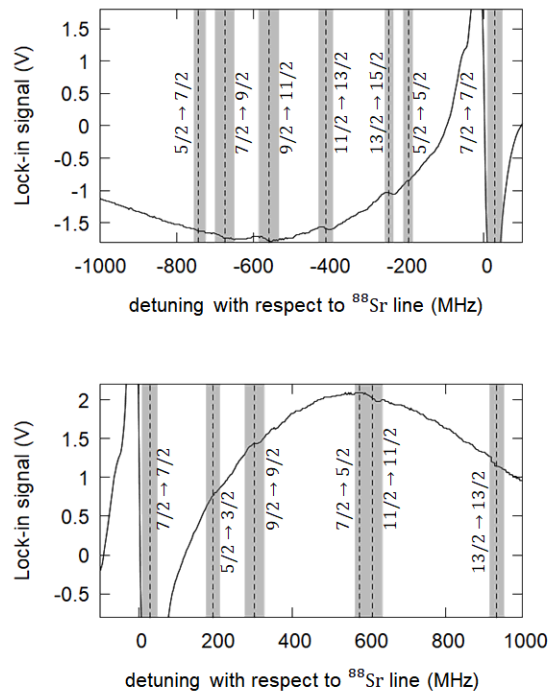}
		\caption{Details of the FMS signal for the hyperfine transitions of ${}^{87}\mathrm{Sr}$. Eight traces were averaged. Dashed lines indicate the calculated positions of the hyperfine transitions ($F\ \rightarrow\ F'$) and gray areas indicate the uncertainty (see text).}
		\label{fig:496_2}
	\end{center}
\end{figure}

\begin{figure}[t]
	\begin{center}
		\includegraphics[width=75mm]{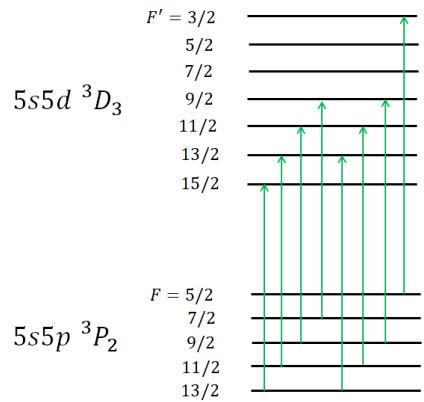}
		\caption{Hyperfine structures of the $(5s5p){}^3P_2$ and the $(5s5d){}^3D_3$ states (not to scale). The solid arrows indicate the transitions observed in this work (see text).}
		\label{fig:hyperfinelevels}
	\end{center}
\end{figure}

\begin{figure}[t]
	\begin{center}
		\includegraphics[width=75mm]{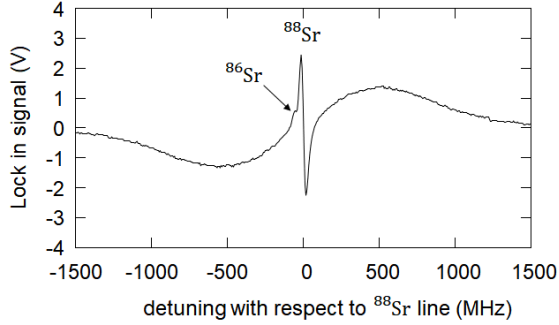}
		\caption{FM spectroscopy signal for the $(5s5p){}^3P_2-(5s5d){}^3D_2$ (497 nm) transition.}
		\label{fig:497}
	\end{center}
\end{figure}

\begin{figure}[t]
	\begin{center}
		\includegraphics[width=75mm]{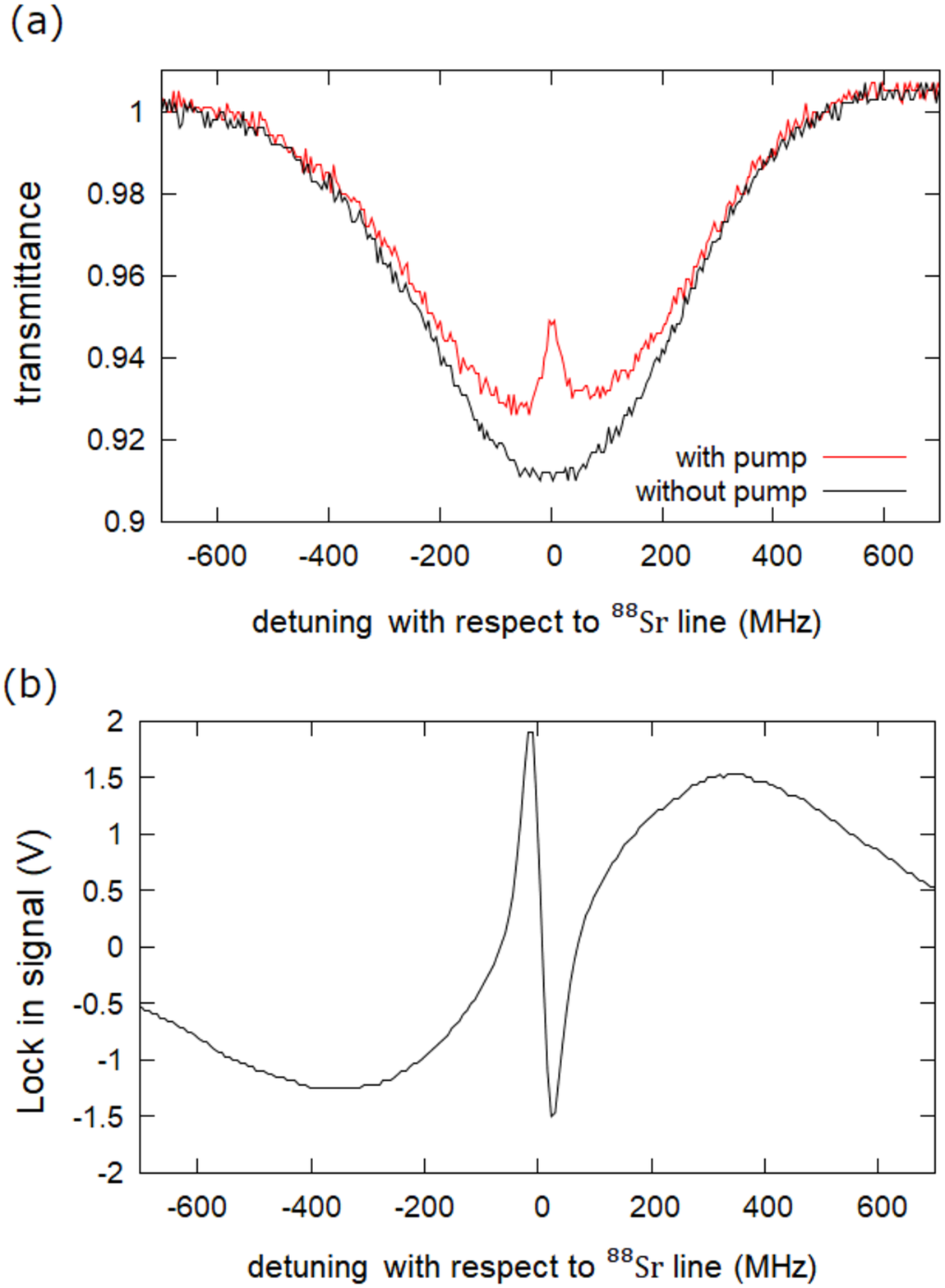}
		\caption{Saturated absorption spectroscopy signal (a) and FM spectroscopy signal (b) for the $(5s5p){}^3P_1-(5s6s){}^3S_1$ (688 nm) transition.}
		\label{fig:688}
	\end{center}
\end{figure}

\begin{figure}[t]
	\begin{center}
		\includegraphics[width=75mm]{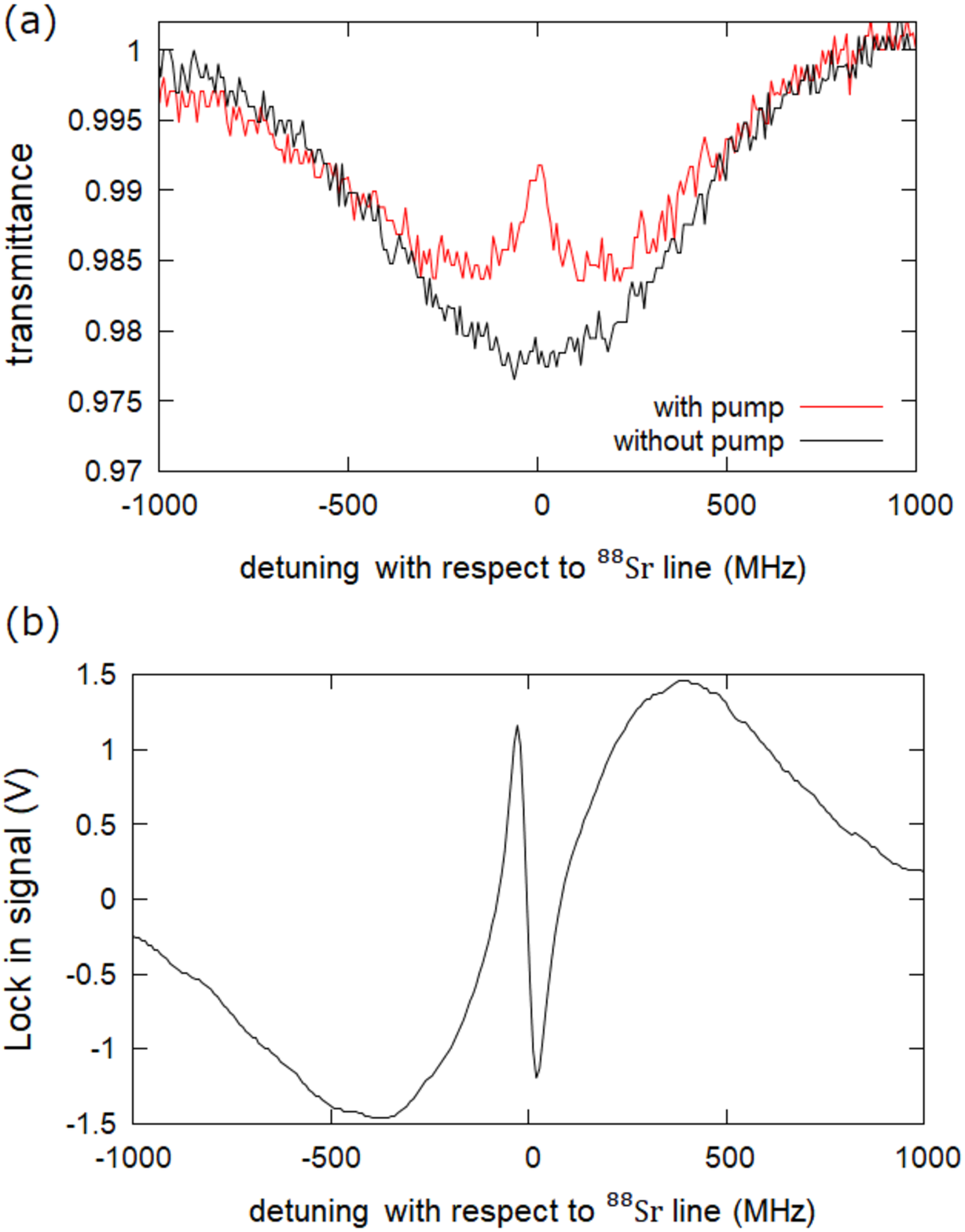}
		\caption{Saturated absorption spectroscopy signal (a) and FM spectroscopy signal (b) for the $(5s5p){}^3P_0-(5s6s){}^3S_1$ (679 nm) transition.}
		\label{fig:679}
	\end{center}
\end{figure}

As shown in Fig. \ref{fig:496_1}, the maximum absorption of the probe beam for the $(5s5p){}^3P_2-(5s5d){}^3D_3$ transition at 496 nm was 20 \% in the custom HCL, while that in the commercial HCL was only 6 \%. When the pump beam was irradiated, a clear Lamb dip was observed in the custom HCL. However, in the commercial HCL, the pump beam merely created a broad Doppler pedestal owing to VCCs \cite{vcc} and a Lamb dip was barely observed.

The FMS signals of this transition for the commercial and custom HCLs are shown in Fig. \ref{fig:496_3}. The amplitude of the FMS signal for the custom HCL was approximately 50 times larger than that for the commercial HCL. This significant improvement of the signal-to-noise ratio led to the observation of the signal from $^{86}\mathrm{Sr}$, which is at $-46.6\ \mathrm{MHz}$ relative to $^{88}\mathrm{Sr}$ \cite{energylevel3}.

The linewidth of the Doppler-free profile for the commercial HCL was 77 MHz, which is much larger the natural linewidth of 10 MHz due to collisional broadening. For the custom HCL, this linewidth was reduced to 27 MHz owing to the reduction of the total buffer-gas pressure.

We also observed the signals from $\mathrm{{}^{87}Sr}$ with a nuclear spin $I=9/2$, as shown in Fig. \ref{fig:496_2}. In a recent work by Stellmer and Schreck \cite{energylevel3}, which used magnetically-trapped $\mathrm{{}^{87}Sr}$ atoms in the $(5s5p){}^3P_2$ metastable state, four of the fifteen hyperfine transitions were observed. They speculatively assigned these transitions to $(F\ \rightarrow\ F')=(13/2\ \rightarrow\ 15/2),\ (13/2\ \rightarrow\ 13/2),\ (13/2\ \rightarrow\ 11/2),\ (5/2\ \rightarrow 3/2)$ and derived the magnetic-dipole constant $A=-156.9\ (3)\ \mathrm{MHz}$, and the electric-quadrapole constant $\ Q=0\ (30)\ \mathrm{MHz}$. According to these values, the frequencies of the hyperfine transitions were calculated as shown in Fig. \ref{fig:496_2} (gray areas indicate uncertainty arising from A and Q). We observed at least eight transitions, six of which were newly observed ($(F\ \rightarrow\ F')=(11/2\ \rightarrow\ 13/2),\ (9/2\ \rightarrow\ 11/2),\ (7/2\ \rightarrow 9/2),\ (9/2\ \rightarrow 9/2),\ (11/2\ \rightarrow 11/2),\ (9/2\ \rightarrow 7/2)$) (see Fig. \ref{fig:hyperfinelevels}). The positions of these transitions were consistent with the calculations, which confirmed that the assignment of the hyperfine transitions in Ref. \cite{energylevel3} were correct. We also determined the hyperfine constants from our own data of all the eight hyperfine transitions as $A\ =\ -157.0\ (3)\ \mathrm{MHz}$, $Q\ =\ -9\ (10)\ \mathrm{MHz}$, which are in good agreement with the values reported in Ref. \cite{energylevel3}. The uncertainties came from frequency jitter of the laser ($\sim 1\ \mathrm{MHz}$) during the scan period (2 s).

Figure \ref{fig:497} shows the FMS signal for the $(5s5p){}^3P_2-(5s5d){}^3D_2$ (497 nm) transition, which is one of the repumping transitions. A clear Doppler-free error signal was also observed. The maximum absorption of the probe beam for this transition was approximately 4 \%, which is one-fifth of that for the $(5s5p){}^3P_2-(5s5d){}^3D_3$ transition. This was consistent with the expected ratio of line strength (100:18) based on the LS-coupling scheme. The signal from $^{86}\mathrm{Sr}$, which was at $-47.5\ \mathrm{MHz}$ relative to $^{88}\mathrm{Sr}$ \cite{energylevel3}, was also observed.

The results of Doppler-free spectroscopy for the $(5s5p){}^3P_1-(5s6s){}^3S_1$ (688 nm) and the $(5s5p){}^3P_0-(5s6s){}^3S_1$ (679 nm) transitions are shown in Fig. \ref{fig:688} and Fig. \ref{fig:679}, respectively. The maximum absorptions of the probe beam were approximately 9 \% for the $(5s5p){}^3P_1-(5s6s){}^3S_1$ transition and 2.3 \% for the $(5s5p){}^3P_0-(5s6s){}^3S_1$ transition. This was also consistent with the expected ratio of line strength (3:1). The signal-to-noise ratios of the error signals for both transitions were sufficient for laser frequency stabilization.

The pressure of the buffer gas in a HCL usually decreases during operation due to adsorption of the buffer gas molecules on the surface within the lamp (gas clean-up), leading to an increase of the sustaining voltage and eventually to the end of the lamp life. We did not observe such increase of the sustaining voltage after an operation period of more than 1000 hours for both the commercial and custom HCLs.

\section{Conclusion}
We demonstrated Doppler-free frequency modulation spectroscopy for metastable Sr atoms. We used a mixed gas of Ne 0.5 Torr and Xe 0.5 Torr as a buffer gas in an HCL to suppress the VCC effect and increase the population of the metastable state of Sr. The maximum absorption of the probe beam for the transitions of $(5s5p){}^3P_2-(5s5d){}^3D_3$ at 496 nm, $(5s5p){}^3P_2-(5s5d){}^3D_2$ at 497 nm, $(5s5p){}^3P_1-(5s6s){}^3S_1$ at 688 nm and $(5s5p){}^3P_0-(5s6s){}^3S_1$ at 679 nm were 20 \%, 4\%, 9\%, and 3\%, respectively. The ratio of absorption was consistent with the theoretical relative line strength based on the LS-coupling scheme. We performed FM spectroscopy for these transitions and obtained error signals with sufficient signal-to-noise ratio for laser frequency stabilization.

For the $(5s5p){}^3P_2-(5s5d){}^3D_3$ transition of fermionic $\mathrm{{}^{87}Sr}$, we observed eight of the fifteen hyperfine transitions and verified the hyperfine constants of the $(5s5d){}^3D_3$ state, which were recently reported \cite{energylevel3}. We are currently trying to further improve the signal-to-noise ratio and reduce the uncertainty of the hyperfine constants.

As the laser frequency can be easily stabilized to the $(5s5p){}^3P_2-(5s5d){}^3D_3$ transition using the HCL investigated in this work, magneto-optical trapping (MOT) using this transition seems to be feasible. However, as reported in Ref. \cite{energylevel3}, this transition is considered inappropriate for MOT owing to the decay from the $(5s5d){}^3D_3$ state to the $(5s5p){}^3P_1$ state via several intermediate states, which leads to the loss of atoms from the cooling cycle. This problem can be avoided by introducing a repumping laser tuned to the $(5s5p){}^3P_1-(5s6s){}^3S_1$ transition at 688 nm. The temperature of MOT using the $(5s5p){}^3P_2-(5s5d){}^3D_3$ transition would be below the Doppler-limit ($\sim 270\ \mu \mathrm{K}$) because the lower level has non-zero angular momentum ($J=2$) and polarization gradient cooling would work.

The method investigated in this work can be readily applied to Doppler-free spectroscopy for the other transitions from the metastable states of Sr atoms and for other alkaline-earth like atoms such as Yb, Cd and Er.

\acknowledgments{
We thank Y. Yamanaka and H. Nagasawa for technical assistance in the development of the laser system. This work was supported by JSPS KAKENHI Grant Number JP23740305, JP17H02881, JP15H02027.}

\bibliography{bib.bib}

\begin{thebibliography}{27}
\bibitem [{wav()}]{wavelengthmeter}%
  \BibitemOpen
  \href@noop {} {\ }\bibinfo {note} {For example, HighFinesse WS8-2 wavemeter
  has an absolute accuracy of 2 MHz.}\BibitemShut {Stop}%
\bibitem [{acc()}]{accurate}%
  \BibitemOpen
  \href@noop {} {\ }\bibinfo {note} {The accurate pressure of the buffer gas is
  not disclosed by the manufacturer.}\BibitemShut {Stop}%
\end{thebibliography}

\end{document}